\documentclass[journal]{IEEEtran}

\ifCLASSINFOpdf
\usepackage[pdftex]{graphicx}
 
\usepackage{amsmath} 

\ifCLASSOPTIONcompsoc
 \usepackage[caption=false,font=normalsize,labelfont=sf,textfont=sf]{subfig}
\else
 \usepackage[caption=false,font=footnotesize]{subfig}
\fi

\newcommand\tab[1][0.4cm]{\hspace*{#1}}
\usepackage{flushend}
\usepackage{tabularx,booktabs,textcomp}
\newcolumntype{C}{>{\centering\arraybackslash}X} 
\newcolumntype{L}{>{\raggedright\arraybackslash}X} 

\usepackage[noadjust]{cite}

\begin{document}
\title{Optical Simulation and Analysis ofGaAs Nanocone, Inverted Nanocone, and Hourglass Shaped Nanoarray Solar Cells}


\author{Sambuddha~Majumder,
       Sooraj~Ravindran}
\maketitle


\begin{abstract}

In this paper, we have investigated the optical properties of GaAs nanocone (NC), inverted nanocone (INC), and hourglass (HG) shaped solar cells by performing Finite Difference Time Domain (FDTD) simulations. The different structures are compared for their light absorptance, reflectance, transmittance, and photogeneration with traditional cylindrical nanowires. The variation of the different optical properties with the sidewall angle is observed for all the structures. It is seen that the best performance of NC, INC, and HG is obtained at a sidewall nanocone angle ($\theta_{b}$) of $\sim$ 7$^{\circ}$, $\sim$ 4$^{\circ}$, and at a sidewall bottom nanocone angle ($\theta_{b}$) of $\sim$ 9$^{\circ}$ respectively, with the NC having the best overall performance followed by INC, except for very large angles where the performance of HG is better than INC. The optical properties of these structures were also observed for different sun angles; the HG and INC structures show improvement in absorption for slightly larger sun angles attributed to their better light trapping properties. Finally, a period study is done to observe the effect of dense and sparse cell packing on optical properties and photogeneration profiles of the nanostructures and it is observed that the sparsely spaced nanowires has a better and more evenly distributed photogeneration profile, which will greatly improve the carrier extraction of the radial junction solar cells.

\end{abstract}




\section{Introduction}
\IEEEPARstart 
{S}{emiconductor} devices and nanowires have become a major frontier of optoelectronic research in the current age with their applications in LED \cite{guo2010catalyst, qian2005core}, LASERS \cite{duan2003single, saxena2013optically, wilhelm2017broadband}, Solar Cells  \cite{garnett2011nanowire, kayes2008radial, hochbaum2010semiconductor, sandhu2014detailed, sun2011compound, lapierre2013iii}, Photodetectors \cite{dai2014GaAs, wang2013high} etc. In particular, a great deal of research is going in cylindrical GaAs nanowire array solar cells mainly for their improved performance over bulk solar cells in photovoltaic applications. Further, nanowire solar cells (NWSC) can be grown on a large range of lattice-mismatched materials because of their smaller diameters \cite{wen2011theoretical}, allowing them to attain efficient multijunction configurations. Also, compared to thin films, the improvement in optical absorptance of NW arrays is attributed to the presence of leaky mode resonances (LMRs) \cite{zhang1998electromagnetic, aw1983optical}. This phenomenon can be used to create wavelength selectivity for
multispectral photodetectors \cite{mokkapati2015optical, azizur2015wavelength} or broadband absorption for solar photovoltaic cells \cite{mokkapati2016review} as the HE$_{1n}$ modes can efficiently couple to vertically standing NWs, resulting in an optical antenna effect \cite{yang1993numerical, fountaine2014resonant, dagyte2018modal}, and produce resonant absorption peaks, which shift to longer wavelengths with increasing NW diameter to satisfy the electromagnetic boundary conditions \cite{wang2012tunable, fountaine2014resonant, dagyte2018modal}. The control of the NW geometry and junction positions are also critical factors that determine the device's performance. Various studies have been done to optimize the radius, period, height, and junction positions of NWSCs \cite{li2015influence, wu2018optimization, huang2012broadband, wen2011theoretical}. Studies have also been done on the different lattice geometries and different cross-sections \cite{mariani2013GaAs, gu2011optical}, and it is seen that this has little effect on optical absorption for large inter-cell separations where there is negligible inter-cell coupling of optical modes. It can also be seen that despite the various methods to optimize the light absorption in the GaAs nanowires, most of the photoabsorption is confined to the top of the nanowires \cite{wu2018optimization, li2015influence}. \\
\tab Therefore, it can be said that constant diameter nanowires with symmetric cylindrical geometries are not the optimal choice to obtain the maximum efficiency configuration. Hence, studies have been done on arrays with multiple diameters \cite{sturmberg2012nanowire, fountaine2016near, yan2018enhanced, lee2020selective, adibzadeh2020optical} and periodic diameter modulation \cite{ko2016periodically, kim2019geometric}. One more design morphology that has emerged is diameter variation along the height of the nanowire which has let to the research in many new tapered nanowire designs. Some of them include the nanocones or frustum nanocones \cite{zhan2014enhanced, wang2014strong, li2016nanostructured, kordrostami2020high, jeong2013all, wang2012enhanced,diedenhofen2011strong, gibson2019tapered, abdel2017electrical, wilson2021simulation, zhang2018photovoltaic}, inverted nanocones \cite{shalev2015enhanced, prajapati2018light, prajapati2018broadband, ko2015high, wilson2021simulation}, hourglass  \cite{seo2020solar} shaped solar cells; due to their improved light absorption, anti-reflective, and light trapping properties. The tapered shape of the nanocones leads to a gradual change in the refractive index profile that improves the anti-reflection properties. For the nanocone arrays with a large slope angle, the filling ratio at the top of the arrays is extremely low, leading to a nearly perfect impedance match with air and almost zero reflection \cite{zhang2018photovoltaic, wu2017efficient}. Also, the expanding diameter improves the broadband absorption of the nanocone due to the broadening of the  HE$_{1n}$ modes \cite{wilson2021simulation}. The inverted nanocone structure has been inspired by the working principle of the human retina \cite{shalev2015enhanced, prajapati2018light, prajapati2018broadband, ko2015high, li2012broadband, marko2019subwavelength}. They can also experience Total Internal reflection (TIR) and  provide enhanced optical absorption compared to NWs due to the support of multiple optical modes. They also lead to an improvement in the trapping of light between them due to their inverted taper shape. The HG structure is a combination of both the nanocones and the inverted nanocones; they consist of upper and lower asymmetric parts, which not only serves as a good anti-reflective layer but also increases the light-trapping path induced by resonant modes. \\
\tab 
Works done by Wilson and Lapierre \cite{wilson2021simulation}, showed the absorption and mode coupling performance of GaAs conical nanowires, they have also cited the different techniques available for the fabrication of nanocones. Zhang $et~al.$ \cite{zhang2018photovoltaic} examined the radial and axial GaAs nanocones for their photovoltaic properties. Fountaine et al. \cite{fountaine2014near} demonstrated that GaAs conical NWs improve absorption over cylindrical NWs, showing an increase in photocurrent from 25.0 mA.cm$^{-2}$ to 29.5 mA.cm$^{-2}$. We have also done photovoltaic studies on the effect on the angle and period of GaAs/AlGaAs nanocones \cite{majumderpv}.  Kordrostrami $et~al.$ \cite{kordrostami2020high} showed photocurrent variation for different Si nanowires and nanocones. Shalev $et~al.$ \cite{shalev2015enhanced} compared absorption enhancements of inverted Si nanocones and cylinders with thin films. Seo $et~al.$ \cite{seo2020solar} demonstrated the influence of Si hourglass-shaped solar cells for improved light trapping path and formation of Whispering Gallery modes. Bai $et~al.$ \cite{bai2014one} showed the influence of the Si nanocone period on its photovoltaic properties.

\tab In this dissertation, three-dimensional (3D) Finite Difference Time Domain (FDTD) simulations are done to observe and compare the optical properties of different GaAs nanocones, inverted nanocones, and hourglass structures and observe their dependence on the nanocone angle and pitch.  In Section~\ref{sec:mm}, the modeling of the structures and the methods used are discussed. Section~\ref{sec:pn}, Section~\ref{sec:pin}  and Section~\ref{sec:phg} examines the various optical properties of nanocones, inverted nanocones, and hourglass respectively; including absorptance, reflectance, transmittance, photogeneration, and ultimate short circuit current for various nanocone angles. The comparative analysis of these designs with each other is done in Section~\ref{sec:dc}. Finally, in Section~\ref{sec:nps}, a study of the solar cell period is done to observe the effect of dense and sparse packing on the optical properties.

\section{Modeling and Methods}
\label{sec:mm}
\begin{figure}[t]
  \centering
  \includegraphics[width=3.5 in]{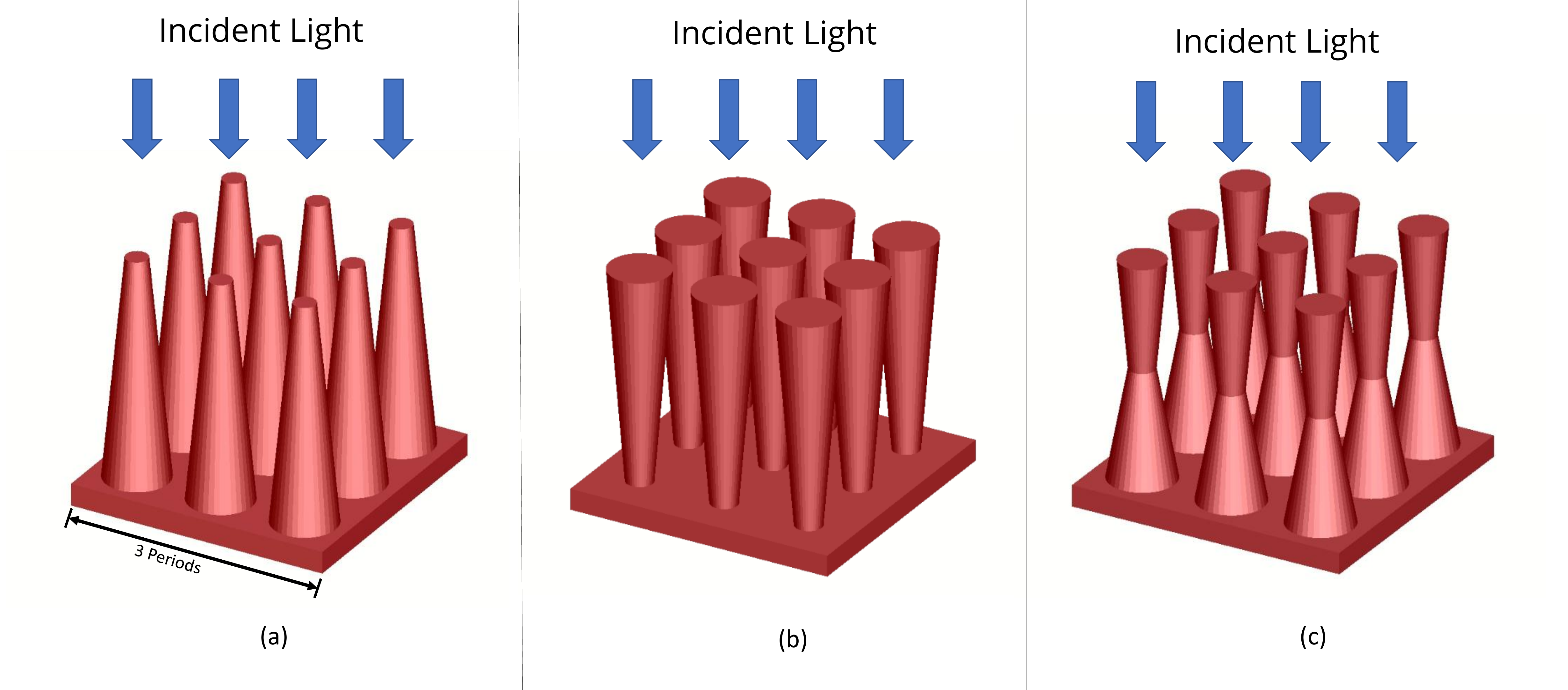}\\
  \caption{Simulation models of (a) Nanocone (b) Inverted nanocone (c) Hourglass solar cell arrays.}
  \label{fig:z}
\end{figure}

In this dissertation, we have modeled GaAs nanocone, inverted nanocone, and hourglass-shaped nanoarray solar cells (Fig~\ref{fig:z}). A unit cell of each of these structures is designed and using periodic boundary conditions; the entire array is simulated. The detailed schematic diagrams for the NCs, INCs, and HGs can be found in Section~\ref{sec:pn}, Section~\ref{sec:pin} and Section~\ref{sec:phg} as Fig~\ref{fig:fa}, Fig~\ref{fig:hc} and Fig~\ref{fig:ja} respectively. A square lattice has been used as the array, \cite{mariani2013GaAs}. The mean diameter and the period is chosen to be 360 nm and 720 nm respectively \cite{zhang2018photovoltaic}. A period study is done again on Section~\ref{sec:nps} to observe the effect of dense and sparse packing on the solar cell performance. It has been shown that for optimal absorption and performance of cylindrical nanowires, the NW diameter is to be taken as 180 nm and D/P = 0.5 \cite{wen2011theoretical}. However, a large diameter is chosen to appreciate the effects of large sidewall tapering angles. The height of each unit cell is taken to be  2 $\mu$m \cite{li2015influence, huang2012broadband} and the height of the substrate is kept at 196 nm \cite{li2015influence}. However, using appropriate PML boundary conditions, it is made to be semi-infinite. \\
\tab A parameter called the `Nanocone angle' is taken for the NC and INC, which is defined as the angle between the sidewall and the normal to the ground surface; it quantifies the changes of the top radius (R$_{TOP}$) and bottom radius (R$_{TOP}$). For HG, two such angles are defined to quantify the taper of the bottom half and the top half as `Nanocone bottom angle ($\theta_{b}$)' and `Nanocone top angle ($\theta_{t}$)' respectively. These angles are varied in each study to observe the different tapering effects.\\
\tab For the optical simulations, the software package `Lumerical FDTD' is used. The absorption per unit volume is calculated using the Poynting vector $\vec{P}$ as:
 \begin{equation}
P a b s=-0.5 \operatorname{real}(\vec{\nabla} \cdot \vec{P})
\end{equation}
Which can be written in a more numerically stable form : 
\begin{equation}
P a b s=0.5 \operatorname{real}\left(i \omega \vec{E} \cdot \vec{D}\right) = 0.5 \omega \varepsilon^{\prime \prime}|\vec{E}|^{2}
\end{equation}
Where, $\varepsilon^{\prime \prime}$ is the imaginary part of permittivity, $\omega$ is the angular frequency of the incident light, and $E$ is the electric field intensity. The Photogeneration rate (assuming that each photon absorbed generates an equal number of electron-hole pairs) can be written as: 
\begin{equation}
G_{p h}=\frac{|\vec{\nabla} \cdot \vec{S}|}{2 \hbar \omega}=\frac{\varepsilon^{ \prime \prime}|\vec{E}|^{2}}{2 \hbar}
\end{equation}
Where $\hbar$ is the reduced Plank's constant, $ G_{p h}$ is weighed by the AM 1.5G solar spectrum and integrated over the entire simulation spectrum. This is obtained using the `Solar Generation' analysis group.  The reflectance and transmittance spectra normalized to the source power are measured using the `Frequency-Domain Field and Power' monitors at the top and bottom of the simulation region respectively. The Quantum efficiency ($\mathrm{QE}(\lambda)$) is defined as :
\begin{equation}
\mathrm{QE}(\lambda)=\frac{P_{a b s}(\lambda)}{\left.P_{i n} (\lambda\right)}
\end{equation}
where, $P_{in}(\lambda)$ and $P_{abs}(\lambda)$ are the powers of the incident light and absorbed light respectively. Assuming that all electron-hole pair contributes to photocurrent, the ultimate short circuit current density Ult-J$_{sc}$ is defined as: 
\begin{equation}
\mathrm{Ult-J}_{s c}=e \int \frac{\lambda}{h c} \mathrm{QE}(\lambda) \mathrm{I}_{\mathrm{AM} 1.5}(\lambda) \mathrm{d} \lambda
\end{equation}
where e is the charge of the electron, $I_{AM1.5}$ is the AM 1.5 solar spectrum, h is the Plank's constant.


\section{Analysis and Results}
\label{sec:ar}


\subsection{Optical Properties of Different Structures}
\label{sec:nas}
Simulations for different optical properties like absorptance, reflectance, and transmittance for the different structures for different angles have been done. The photogeneration and field profiles across the structures have also been observed. Based on the study, a comparison between the different structures is made, and the optimum nanocone angles are obtained.


\subsubsection{Properties of Nanocones}
\label{sec:pn}

\begin{figure}[t]
        \centering
        \subfloat[]{\label{fig:fb}\includegraphics[width=3.45in]{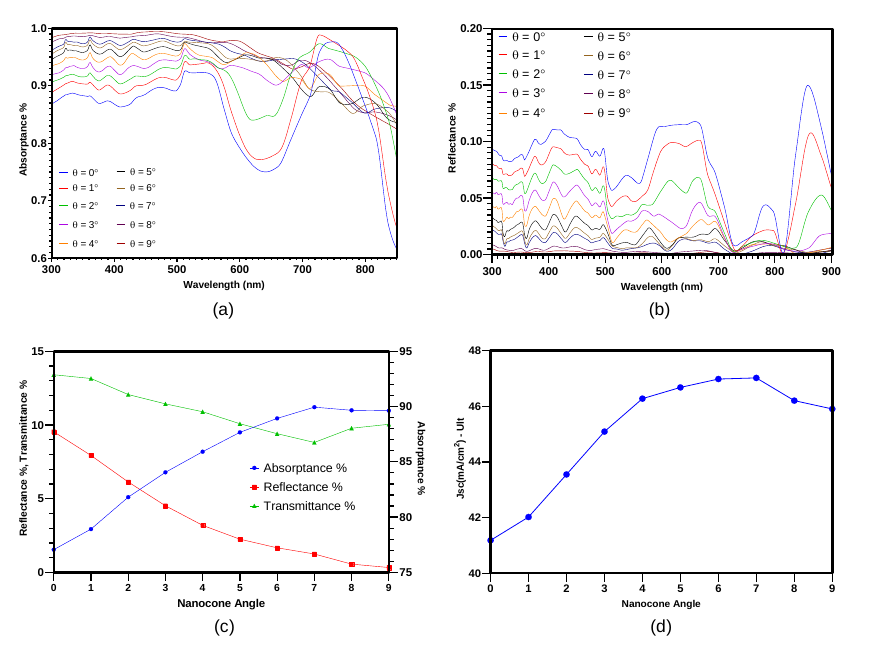}}
        
        \qquad
        \subfloat[]{\label{fig:fa}\includegraphics[width=3.45in]{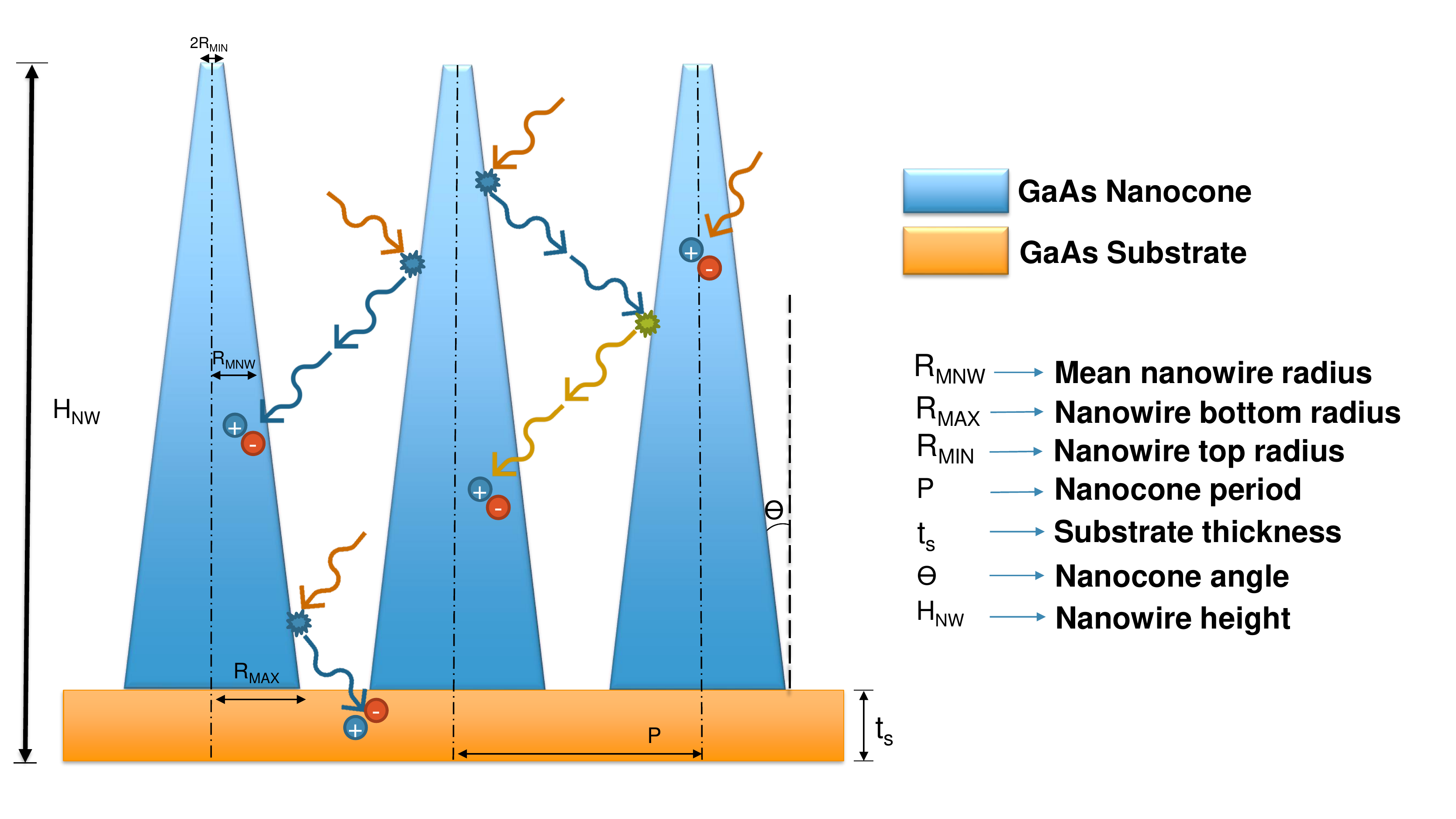}}
        \caption{\textbf{(i)} The (a) Absorptance spectra, (b) Reflectance spectra, (c) Integrated reflectance, transmittance and absorptance, over the entire spectrum, (d) Ultimate J$_{sc}$ of the nanocone solar cells for different $\theta$. \textbf{(ii)} Schematic diagram of the nanocone solar cell array. }
        \label{fig:f}
\end{figure}


\begin{figure}[t]
        \centering
        \subfloat[]{\label{fig:ga}\includegraphics[width=3.45in]{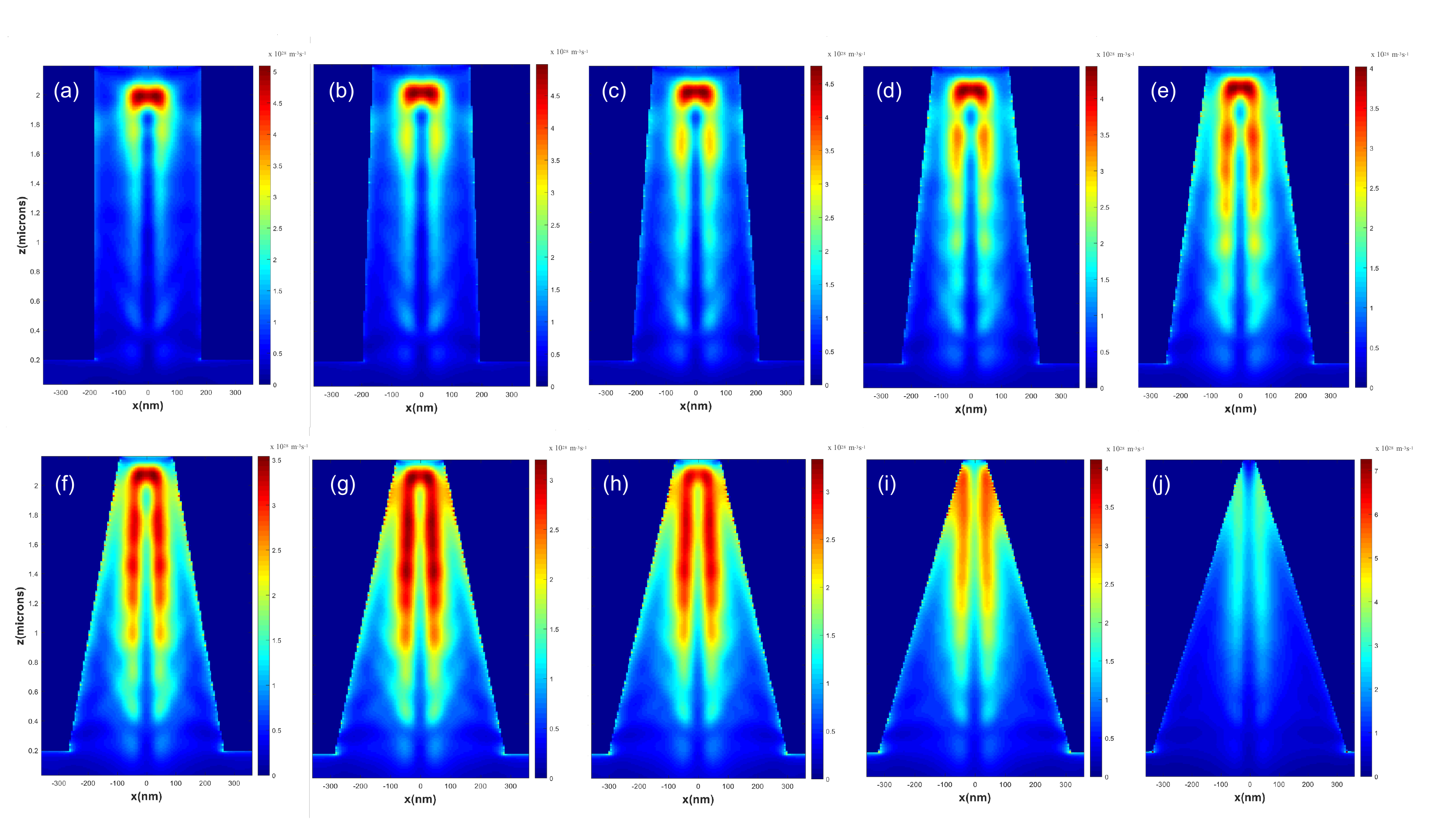}}
        \qquad
        \subfloat[]{\label{fig:gb}\includegraphics[width=3.4in]{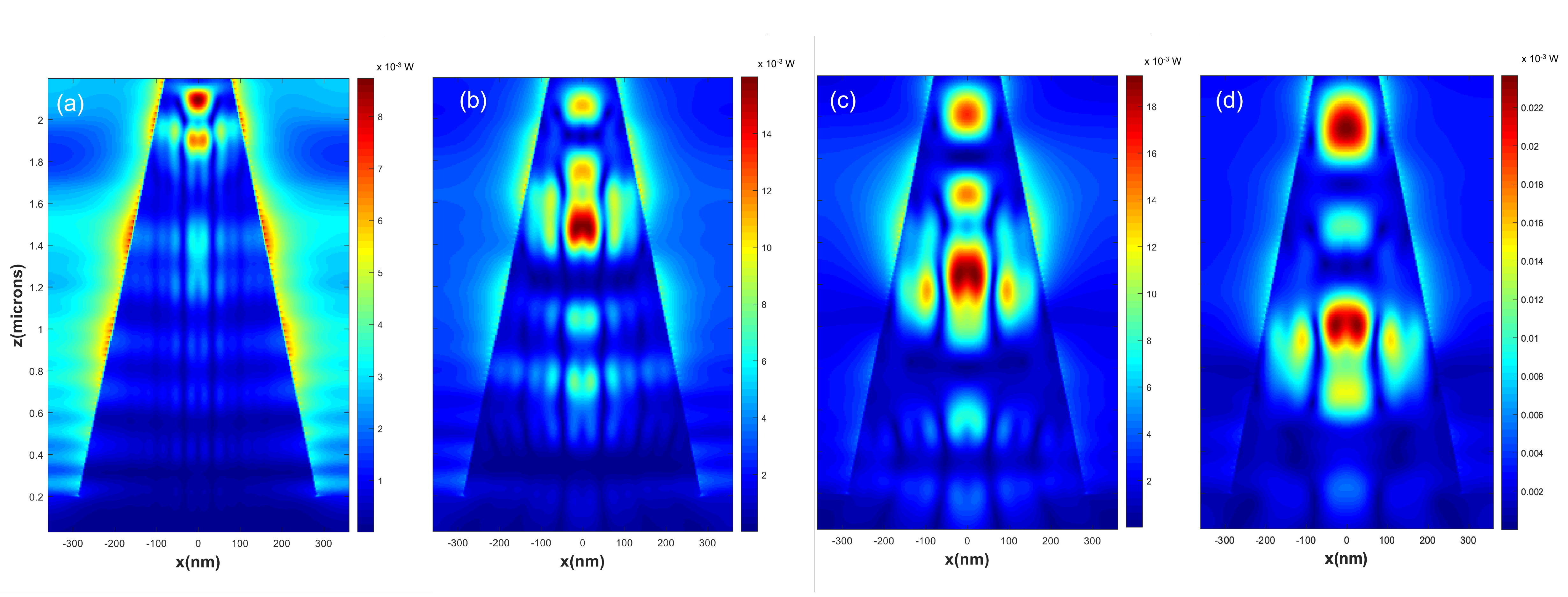}}
        \caption{\textbf{(i)} The photogeneration profiles of nanocone for (a) $\theta$ = 0$^{\circ}$, (b) $\theta$ = 1$^{\circ}$, (c) $\theta$ = 2$^{\circ}$, (d) $\theta$ = 3$^{\circ}$, (e) $\theta$ = 4$^{\circ}$, (f) $\theta$ = 5$^{\circ}$, (g) $\theta$ = 6$^{\circ}$, (h) $\theta$ = 7$^{\circ}$, (i) $\theta$ = 8$^{\circ}$, (j) $\theta$ = 9$^{\circ}$. \textbf{(ii)} The field power profiles across the nanocone for wavelengths: (a) 500 nm, (b) 600 nm, (c) 700 nm, (d) 800 nm. }
        \label{fig:g}
\end{figure}

In this study, the nanocone structure is studied. Initially, a cylindrical nanowire of radius = 180 nm is taken. A tapering is given to its sidewall by varying the top and bottom diameters but keeping the mean nanowire radius (R$_{MNW}$) constant at 180 nm. The nanocone angle ($\theta$) is varied from 0 to 9 degrees, and the changes in the optical properties are observed.\\
\tab From Fig~\ref{fig:fa}-(a), (b), we see the Absorptance and Reflectance spectra for the nanocones for different nanocone angles. We can see that the absorption increases for lower wavelength modes ($\sim$  300 - 700 nm) as the nanocone angle increases. However, for large angles, the absorptance tends to decrease at the large wavelength regime($\geq$ 700 nm). The reflectance decreases throughout the entire wavelength range as the nanocone angle increases. If we look at the overall absorptance, reflectance, and transmittance by integrating them over the entire spectrum (Fig~\ref{fig:fa}-(c)), we see that, with the increase in the nanocone angle, the absorptance increases rapidly up to an angle of 7$^{\circ}$ and then it becomes constant with slight decrease; the reflectance and transmittance decrease with a small increase in reflectance after 7$^{\circ}$. The corresponding ultimate short circuit current density (Ult- J$_{sc}$) can be seen on Fig~\ref{fig:fa}-(d), it follows the same trend as the absorptance and increases rapidly up to an angle of 7$^{\circ}$ followed by a decrease.  \\
\tab The reason for such a behavior of the nanocones can be attributed to two primary factors - improved and extended photoabsorption and sidewall reflection. The anti-reflection ability of the NC arrays can be attributed to the low filling ratio, which reduces the effective refractive index and offers a good impedance match betweenGaAs and air \cite{zhang2018photovoltaic, wu2017efficient}. For the nanocone arrays with a large slope angle, the filling ratio at the top of the arrays is extremely low, leading to a nearly perfect impedance match with air and almost zero reflection. For symmetric cylindrical nanowires, there is a lot of light that is lost due to reflection in the space between the adjacent nanowire cells and also at the top of the nanowire; having a tapered structure reduces that drastically (Fig~\ref{fig:fb}). In a tapered nanocone structure, the space between two adjacent cells is decreased due to an increase in the bottom radius. The reflection from the top surface of the nanowire is also decreased due to its smaller radius. Therefore, more light now falls directly on the sidewall of the structure, and the light reflected off the sidewalls can go and be absorbed into or re-reflected off the adjacent cell, thereby reducing the amount of light lost due to reflection and also increase in the absorptance.\\
\tab From figure (Fig~\ref{fig:ga}), we can see the photogeneration profiles across the nanocone structure for different nanocone angles; it is observed that, when the angle of the nanocone is less, the photogeneration hotspot is mostly confined to the top of the nanowire. By introducing a tapered structure, we observe, as the nanocone angle increases, the photon absorption shifts downward, leading to an enhanced absorption throughout the nanowire. However, after a certain angle ($\sim$ 7$^{\circ}$), we lose the photogeneration hotspot present in the top entirely, and there is a decrease in the photogeneration with further increase in angle; this is also reflected in Fig~\ref{fig:fa}-(c), (d). The reason for such a behavior is that the light absorption in NWs is dominated by resonant modes, which are very closely related to the NW diameter \cite{zhang2018photovoltaic, anttu2010coupling, zhan2014enhanced}. In nanocones, the diameter of the NW continually changes across the nanowire height, with the top diameter being very smaller than the bottom diameter. Due to this unique geometry, only a few long-wavelength modes can be supported in the top, and absorption (particular for long wavelengths) happens towards the thicker middle regions of the structure. From Fig~\ref{fig:gb} we can see the field profile inside the nanocone with an angle of 6$^{\circ}$ for different wavelengths. We see that as the wavelength increases, the absorption shifts to the lower section of the nanocone, and thus there is an even distribution of light absorption throughout the structure for the entire spectrum, which results in the corresponding photogeneration. For large angles, the nanocone top becomes too thin, and the reduction in absorption in the top regions is more dominant than the gain in the middle region, thereby resulting in a decrease in absorptance in the long-wavelength regime. For radial pin junction nanocones, as the absorption shifts downward, the effective absorption length increases, and as the i-region exists in the radial direction, this downward shift leads to an enhanced overlap between the i-region and photogeneration. Therefore, the effective absorption is also believed to be increased along with improvement in carrier extraction \cite{zhang2018photovoltaic, majumderpv}.


\subsubsection{Properties of Inverted Nanocones}
\label{sec:pin}

\begin{figure}[t]
        \centering
        \subfloat[]{\label{fig:ha}\includegraphics[width=3.45in]{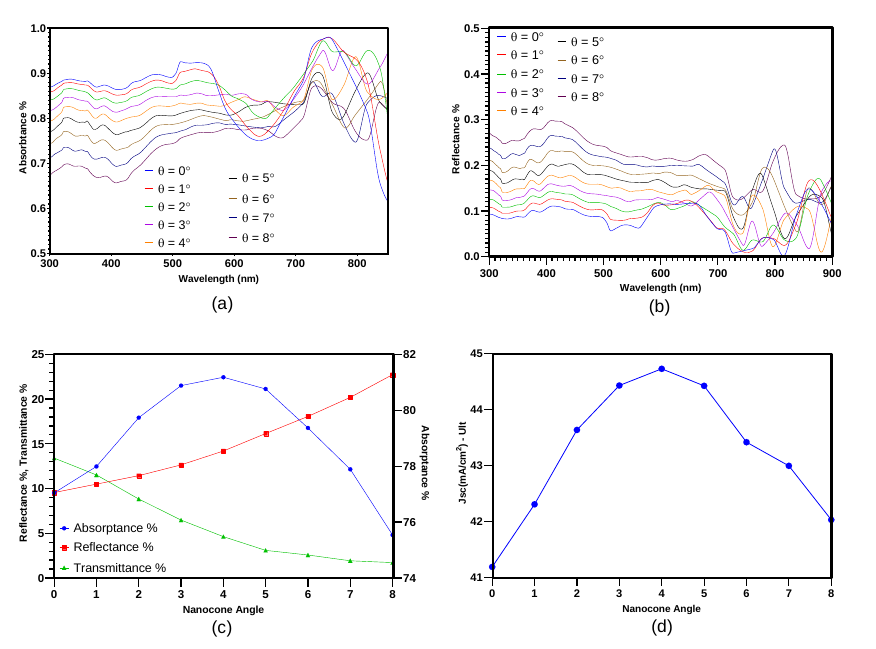}}
        \qquad
        \subfloat[]{\label{fig:hb}\includegraphics[width=3.5in]{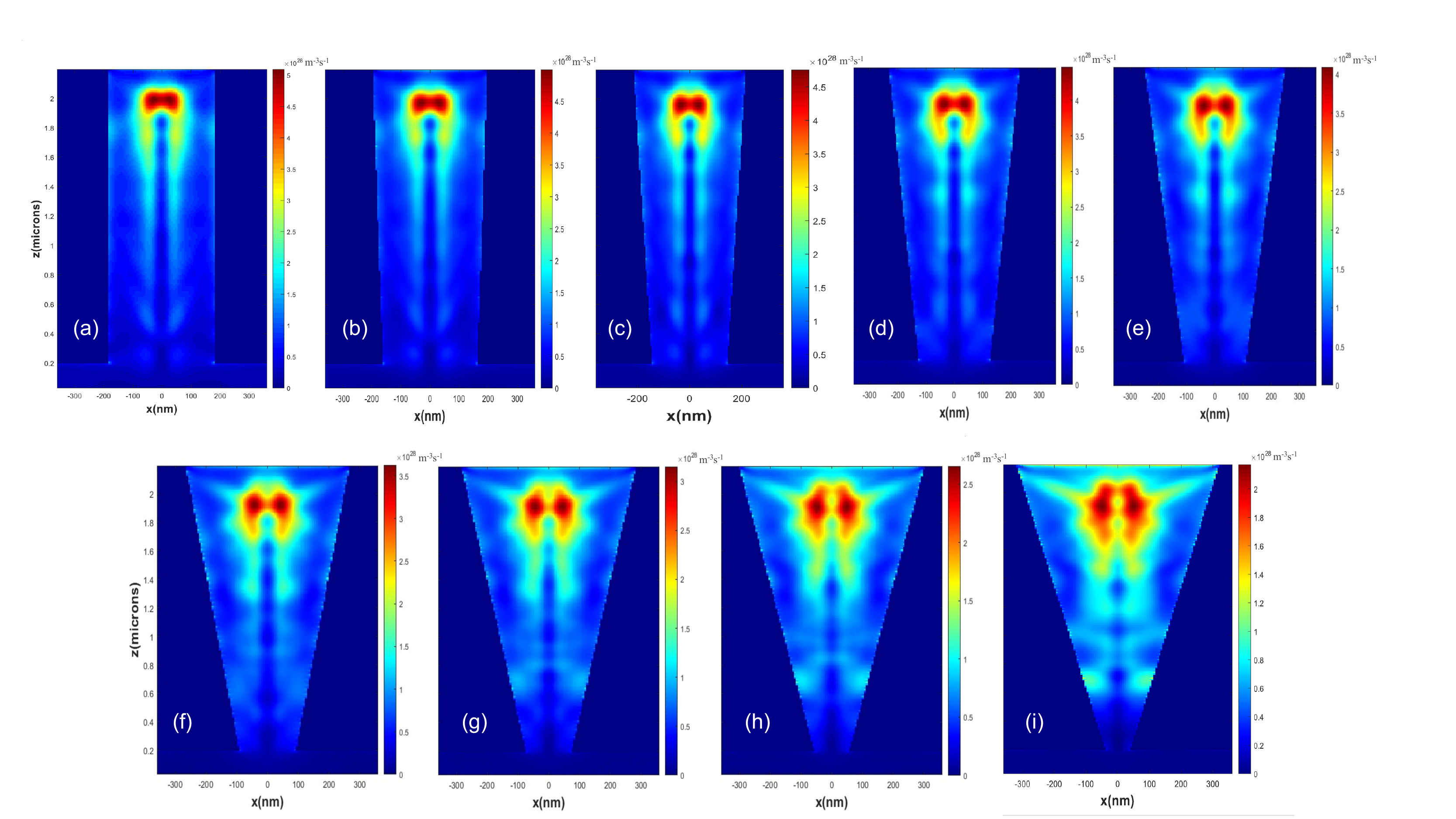}}

        \caption{\textbf{(i)} The (a) Absorptance spectra, (b) Reflectance spectra, (c) Integrated reflectance, transmittance and absorptance, over the entire spectrum, (d) Ultimate J$_{sc}$ of inverted nanocone solar cell array for different $\theta$. \textbf{(ii)} The photogeneration profiles of inverted nanocone for: (a) $\theta$ = 0$^{\circ}$, (b) $\theta$ = 1$^{\circ}$, (c) $\theta$ = 2$^{\circ}$, (d) $\theta$ = 3$^{\circ}$, (e) $\theta$ = 4$^{\circ}$, (f) $\theta$ = 5$^{\circ}$, (g) $\theta$ = 6$^{\circ}$, (h) $\theta$ = 7$^{\circ}$, (i) $\theta$ = 8$^{\circ}$. }
        \label{fig:h}
\end{figure}


\begin{figure}[t]
  \centering
  \includegraphics[width=3.5 in]{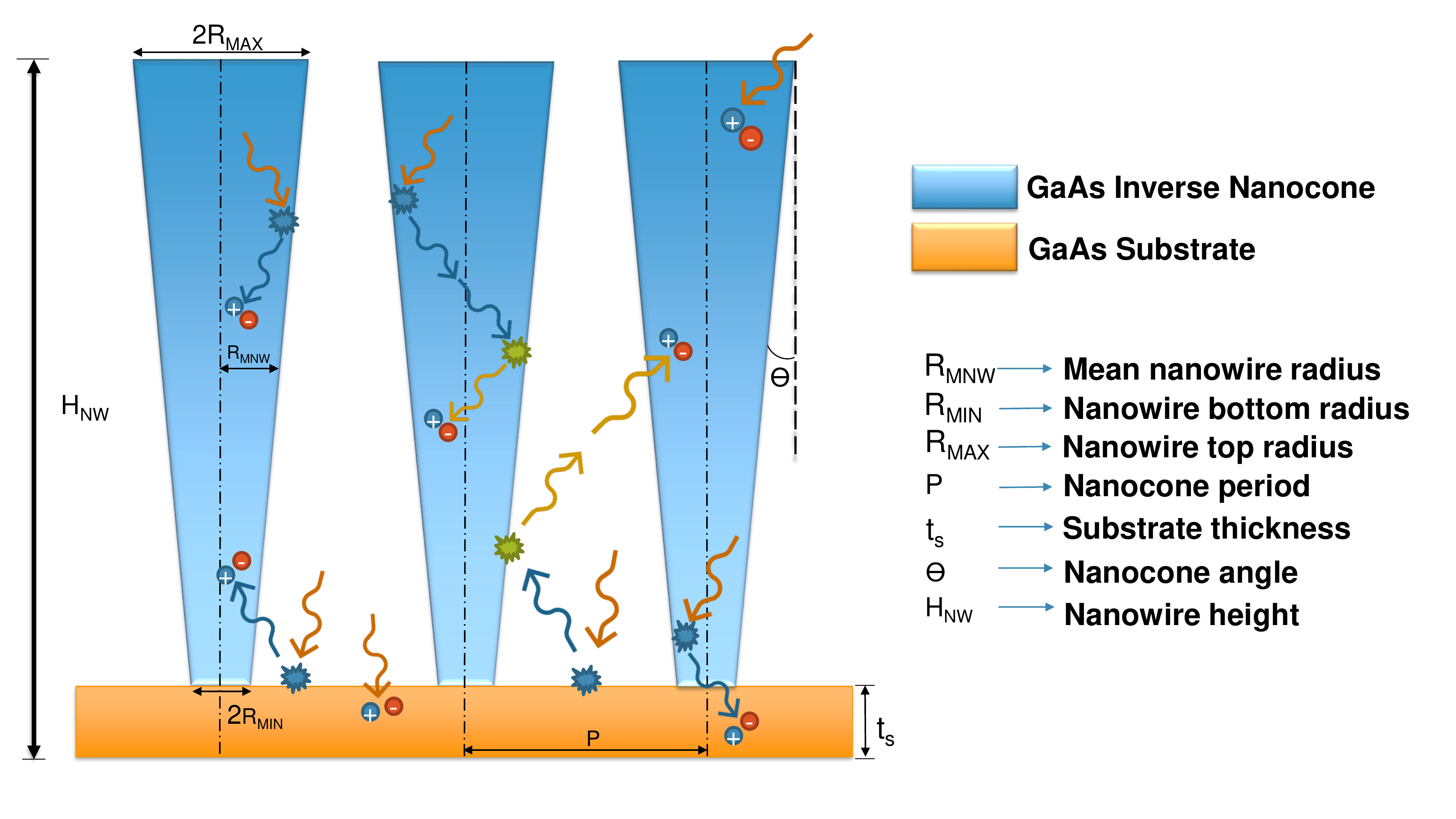}\\
  \caption{Schematic diagram of the inverted nanocone solar cell array.}
  \label{fig:hc}
\end{figure}

In contrast to the previous design, now the nanocone is inverted with a larger top radius and a narrow bottom. The nanocone angle is defined and varied in the same way as the previous case, with the mean radius is kept constant at 180 nm. \\
\tab From Fig~\ref{fig:ha}-(a), (b), we see the Absorptance and Reflectance spectra of the inverted nanocone for different nanocone angles. It can be seen that, as the nanocone angle increases the absorptance in the smaller wavelength regions decreases while the absorptance increases in the larger wavelength range; however, with large angles, the absorption decreases in the large wavelength regime as well. On the other hand, as the angle increases, the reflectance increases for both lower and upper wavelengths. If we look at the overall absorptance, reflectance, and transmittance (Fig~\ref{fig:ha}-(c)), we see that with the increase in angle, the reflectance and transmittance continually increases and decreases respectively, however, the absorption increases and reaches a maximum value at $\sim$ 4$^{\circ}$ and then decreases with further increase in angle and the same is reflected in the curve of The Ult-J$_{sc}$ (Fig~\ref{fig:ha}-(d)). If we look into the photogeneration profiles of these inverted nanocone structures (Fig~\ref{fig:hb}), we see that with the increase in angle, the photogeneration magnitude decreases but is more spread-out throughout the structure. When working with radial junction solar cells, this property can be useful only up to some extent as having a very diffused photogeneration will only lead to large recombination, whereas recombination along the junction will allow carrier separation and subsequent extraction.\\
\tab The reason for such a change in the optical properties is external top and sidewall reflection and total internal reflection. This cell exhibits higher J$_{sc}$ than the symmetric NW solar cells stemming from the increase in the light trapping path leading to improved absorption and the charge carrier generation in the radial junction due to total internal reflection at the sidewall of the asymmetric NW \cite{ko2015high}. Up to an angle of $\sim$  4$^{\circ}$, the structure exhibits TIR for normally incident rays, and it slowly decreases as the sidewall slope increases further. Also, compared to the cylindrical nanowires, there is an improved light trapping between the inverted nanocones, the light that does not hit the nanocones but enters the space between them gets absorbed in the substrate or are reflected from the substrate to the nanocone sidewall where they can be absorbed or re-reflected. A schematic diagram for this behavior is shown in Fig~\ref{fig:hc}. The reason for the increase in reflectance with the increase in angle is because of the wide top radius of the structure, which results in more loss of light due to external reflection, and for large angles ($\geq \sim 4^{\circ}$) the decrease in TIR coupled with increased external reflection due to large R$_{TOP}$ reduces the performance; thus this cell exhibits an interplay of external and internal reflection to obtain the best performance. However, The external reflection can be further reduced using anti-reflection coatings or corrugated surfaces.


\subsubsection{Properties of Hourglass}
\label{sec:phg}

\begin{figure}[t]
        \centering
        \subfloat[]{\label{fig:ia}\includegraphics[width=3.45in]{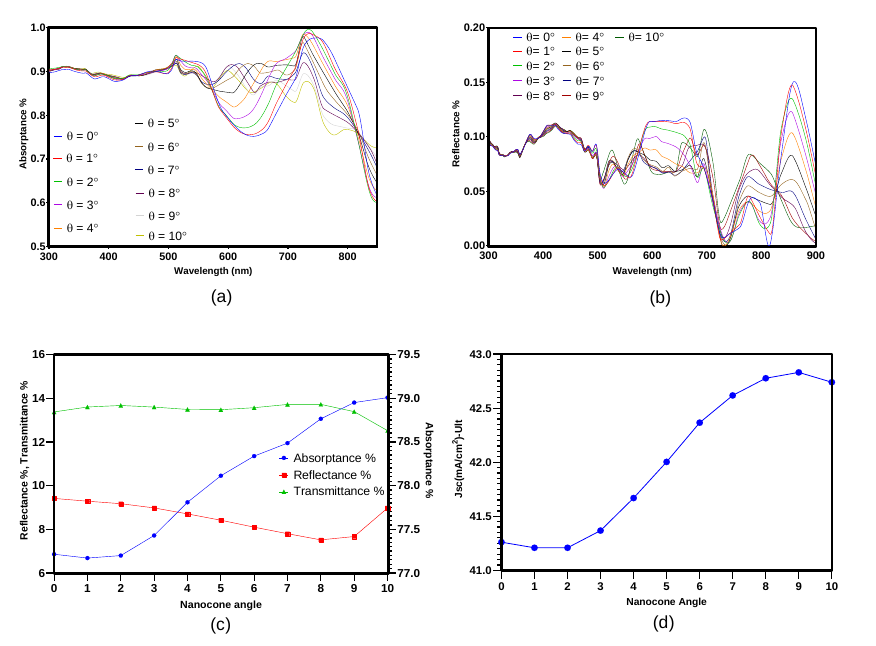}}
        \qquad
        \subfloat[]{\label{fig:ib}\includegraphics[width=3.45in]{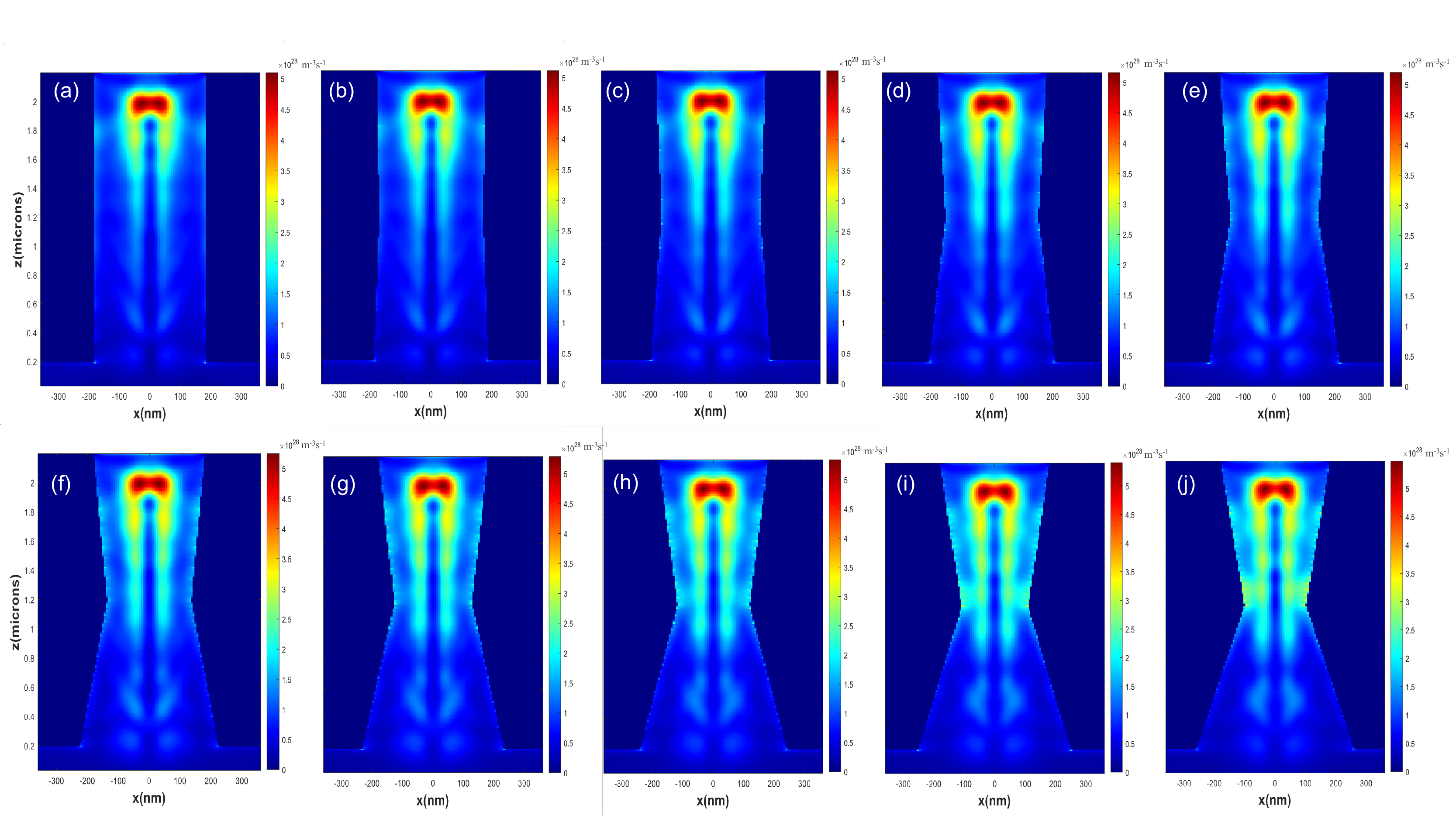}}
         \qquad
        \subfloat[]{\label{fig:ic}\includegraphics[width=3.4in]{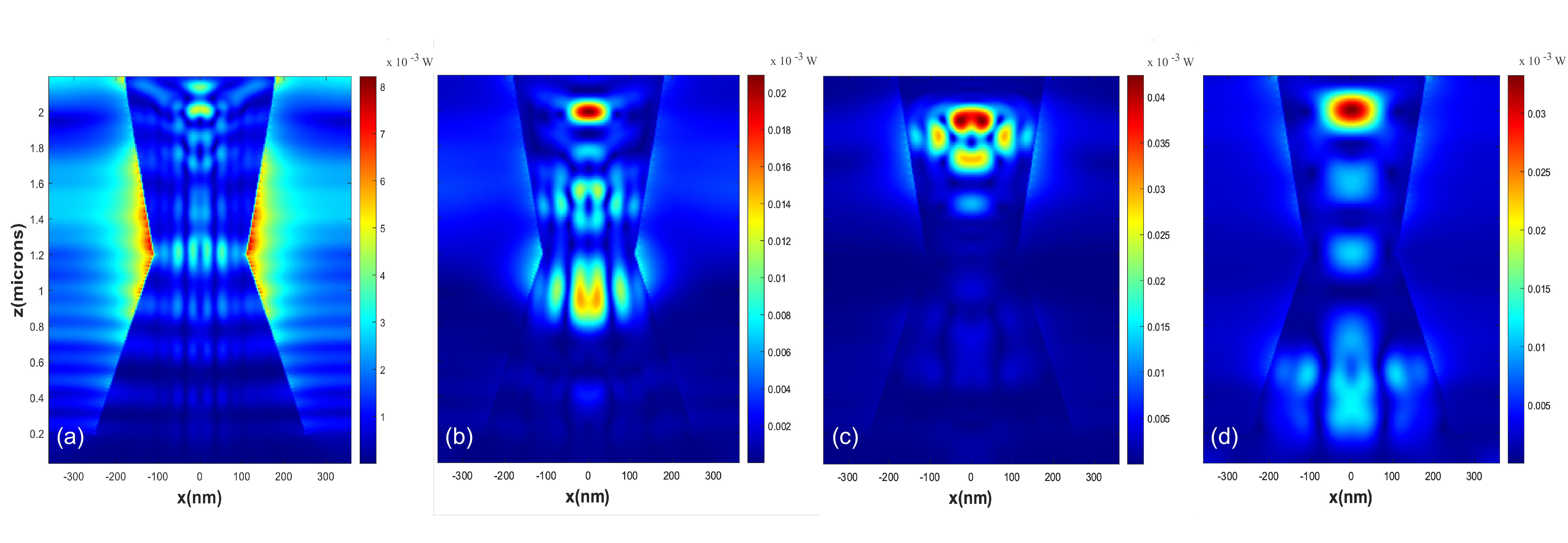}}

        \caption{\textbf{(i)} The (a) Absorptance spectra, (b) Reflectance spectra, (c) Integrated reflectance, transmittance and absorptance, over the entire spectrum, (d) Ultimate J$_{sc}$ of hourglass solar cell array for different $\theta_{b}$. \textbf{(ii)} The photogeneration profiles of hourglass for: (a) $\theta$ = 0$^{\circ}$, (b) $\theta$ = 1$^{\circ}$, (c) $\theta$ = 2$^{\circ}$, (d) $\theta$ = 3$^{\circ}$, (e) $\theta$ = 4$^{\circ}$, (f) $\theta$ = 5$^{\circ}$, (g) $\theta$ = 6$^{\circ}$, (h) $\theta$ = 7$^{\circ}$, (i) $\theta$ = 8$^{\circ}$, (j) $\theta$ = 9$^{\circ}$. \textbf{(iii)} The field power profiles across the hourglass for wavelengths: (a) 500 nm, (b) 600 nm, (c) 700 nm, (d) 800 nm. }
        \label{fig:i}
\end{figure}


\begin{figure}[t]
        \centering
         \subfloat[]{\label{fig:ja}\includegraphics[width=3.45in]{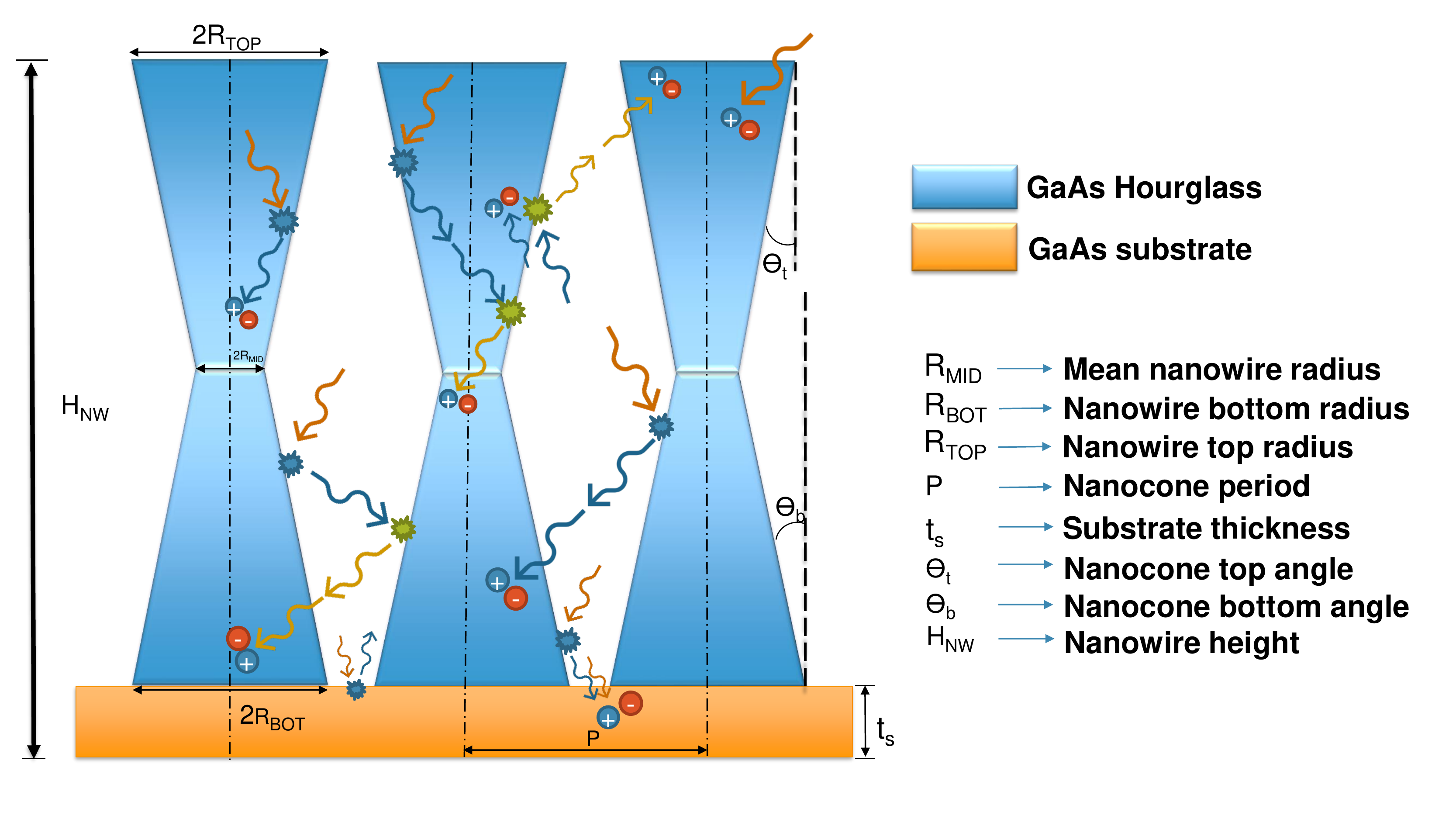}}
           \qquad
        \subfloat[]{\label{fig:jb}\includegraphics[width=3.4in]{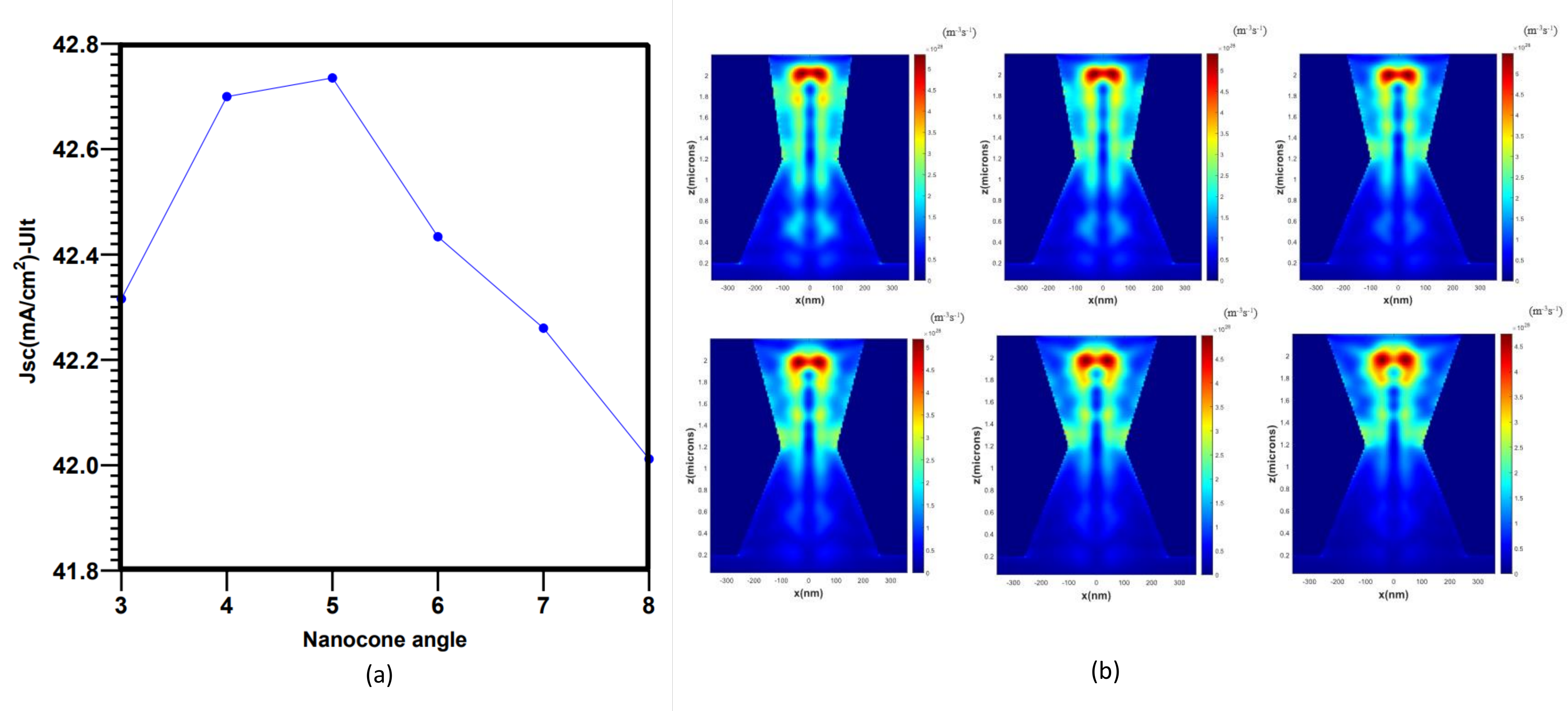}}

        \caption{\textbf{(i)} Schematic diagram of the hourglass solar cell array. \textbf{(ii)} (a) Ultimate J$_{sc}$, (b) Photogeneration profiles of hourglass solar cell for different $\theta_{t}$,  }
        \label{fig:j}
\end{figure}

The Hourglass structure exhibits the properties of both the nanocone and the inverted nanocone solar cell. Initially, the top radius (R$_{TOP}$) is fixed, and the angle of the bottom half defined as `Nanocone bottom angle', $\theta_{b}$ ( will also be referred to as the `Nanocone Angle') of the nanowire; is changed from 0 to 10 degrees by varying the middle radius (R$_{MID}$) and bottom radius (R$_{BOT}$). This also automatically induces a taper in the top half as well, denoted by the `Nanocone top angle' ($\theta_{t}$). It can be seen that as the angle increases, the lower half becomes a nanocone structure while the top attains an inverted nanocone structure.\\
\tab From Fig~\ref{fig:ia}-(a), (b), we see the Absorptance and Reflectance spectra of the hourglass for different nanocone angles. We see that as the angle $\theta_{b}$ increases, there is an improved absorption in some regions of the spectra for small angles, which again decreases for larger angles and vice versa, a similar but more uniform decrease is observed for the reflectance spectra. From Fig~\ref{fig:ia}-(c), which gives the net absorptance, reflectance, and transmittance, we see that the absorption slowly increases until it reaches a maximum at $\sim$ 9 degrees. There is a very consistent transmission with a slight decrease at 9 degrees. While the reflectance gradually decreases with an increase towards the end. The Ult-J$_{sc}$ in Fig~\ref{fig:ia}-(d) accordingly varies in the same way with the maximum at 9$^{\circ}$. The photogeneration plots across the structure for different Nanocone bottom angles are shown in Fig~\ref{fig:ib}. We see that, as the nanocone angle increases, light absorption spreads out along the length of the structure up to some angle, thus increasing the effective absorption length, and with further increase in angle, the photogeneration spreads out more along the width of the structure as well; however, there is no decrease in the magnitude of photogeneration in contrast to what was observed in the case of the nanocone and inverted nanocone structures. From Fig~\ref{fig:gb}, we can see the field profile inside the nanocone with an angle of 9$^{\circ}$ for different wavelengths. We see that for longer wavelengths modes, the field is mostly confined to the wider top region, whereas for smaller wavelength modes, the field exists in the narrow middle region as well; the same principle as the case of nanocones applies here as well, that nanowires are dominated by resonant modes, which are very closely related to the NW diameter. Therefore, the narrow middle region allows for the absorption and photogeneration by the smaller modes while the large modes are absorbed and subsequently lead to the photogeneration of carriers in the wider top region. 
  \\
\tab With the increase in angle $\theta_{b}$, the upper half starts to exhibit TIR similar to the case of inverted nanocone structure. However, unlike pure INC solar cell, the refection does not increase because here, the R$_{TOP}$ is constant. The light trapped between the structures is getting reflected due to the tapering of the bottom half (Fig~\ref{fig:ja}), similar to the nanocone structure. This effect decreases total reflection losses of the cell due to increasing light trapping length. It is observed that when the $\theta_{b}$ is given an angle of 9 degrees, the top half receives an angle $\theta_{t}$ of $\sim$ 4.5 degrees, which is where we are expected to see the maximal absorption for inverted nanocone structures. To further verify this point in Fig~\ref{fig:jb} we have even varied the $\theta_{t}$ from 3$^{\circ}$ - 8$^{\circ}$ keeping $\theta_{b}$ constant and found that the light absorption and the Ult-Jsc are maximum for the angle between 4$^{\circ}$ - 5$^{\circ}$. Further increase in the nanocone angles will lead to loss of TIR, decreased light trapping, and increased losses due to external reflection. Therefore, considering all these factors, an optimal optical design is reached.


\subsubsection{Design Comparison}
\label{sec:dc}

\begin{figure}[t]
  \centering
  \includegraphics[width=3.5 in]{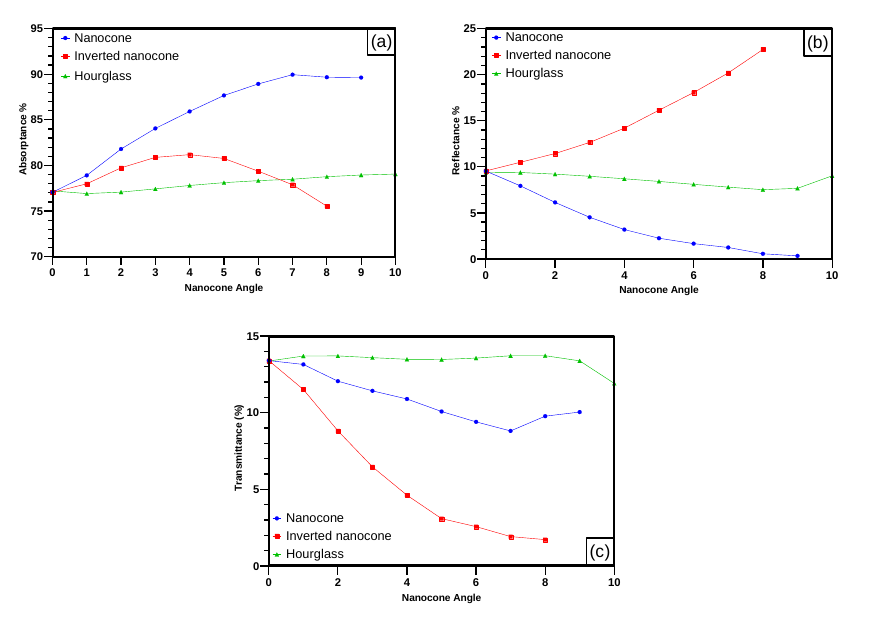}\\
  \caption{Integrated (a) Absorptance, (b) Reflectance, (c) Transmittance of the different solar cell structures for different nanocone angles.}
  \label{fig:k}
\end{figure}
\begin{figure}[t]
  \centering
  \includegraphics[width=3.5 in]{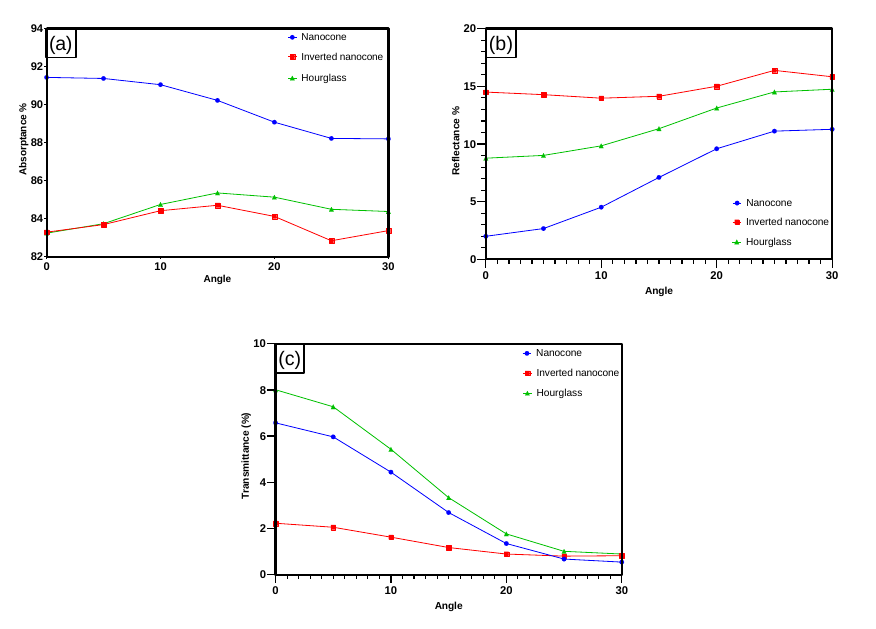}\\
  \caption{Integrated (a) Absorptance, (b) Reflectance, (c) Transmittance of the different solar cell structures for different sun angles.}
  \label{fig:kk}
\end{figure}
In Fig~\ref{fig:k}, we compare these properties of all the designs. We see that, in terms of light absorption and reflection, the nanocone structure has the best overall performance. The INC structure has better absorption than the HG structure up to $\sim$ 7 degrees. The reflectance of NC and HG decreases gradually with nanocone angle, with a larger rate of decrease for the nanocones; however, for INC, the reflectance increases and is much larger than both the other structures. On the other hand, the transmittance of the INC structure decreases rapidly with angle and is much lower than other structures with HG at the top with almost constant transmittance. The HG structure thus has the most tolerance to its sidewall angle and has a steady performance overall.  Individually the best performance of NC, INC, and HG are obtained at a $\theta$ of $\sim$ 7$^{\circ}$, $\sim$ 4$^{\circ}$, and at a $\theta_{b}$ of $\sim$ 9$^{\circ}$ respectively. \\

\tab The net absorptance, reflectance and transmittance of these structures is observed for different sun angles (measured from the normal to the surface). In Fig~\ref{fig:k}, the different properties are measured for sun angle variation from 0 to 30 degrees. It is observed that, as the sun angle increases, The absorptance of the NC decreases continuously whereas, the absorptance of the INC and HG initially increases and then it starts to decrease; and the HG shows a better performance than then INC for greater angles. There is a huge drop in the transmittance of NC and HG with the increase in sun angle, however, the INC structure has very consistent transmittance. Similarly, the reflectance of the INC is almost constant with slight increase for large angles; whereas, the reflectance of the INC and HG increases significantly. \\
\tab This type of behaviour can be linked to the better light trapping properties of the INC and HG structures. When the sun angle increases, the light trapped the the inter-cellular space increases due to inverted tapered sidewall regions resulting in more absorption. With large increase in angles, the reflection loss from the sidewalls increase and the absorptance decreases. However, the NC still has a better performance the the other structures.


\subsection{Period Study}
\label{sec:nps}


\begin{figure}[t]
  \centering
  \includegraphics[width=3.5 in]{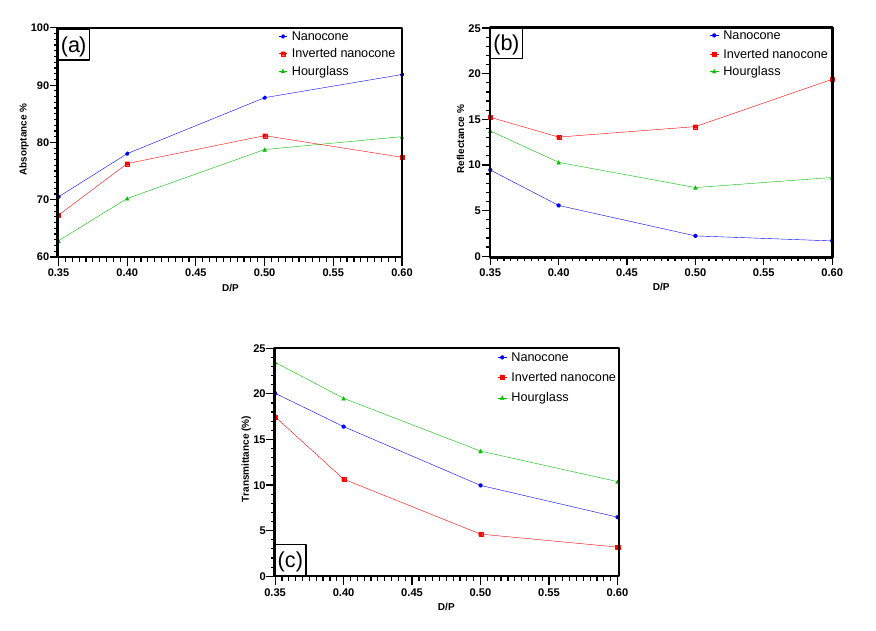}\\
  \caption{Integrated (a) Absorptance, (b) Reflectance, (c) Transmittance of the different solar cell structures for different D/P ratios.}
  \label{fig:m}
\end{figure}

When we are working with symmetric cylindrical nanowires, the packing of the solar cells does not have a very large significance as there is a very little reflection from the sidewalls, and most of the light that is normally incident on the solar cell array is either directly absorbed in the solar cell or are reflected off the space between the nanowires. However, for the case of tapered surfaces, the space between individual cells plays a significant role in determining the light trapping, reflection, absorption, and photogeneration. Therefore, in this section, we observe how the properties of the different structures vary with the change in the period. \\
\tab In Fig~\ref{fig:m}, we see how the absorptance, reflectance, and transmittance of the different structures vary with the increase in the D/P ratio (diameter constant, decrease in period). It can be seen for NC and HG that the absorptance increases while reflectance and transmittance decrease with the decrease in the nanocone period. This is because, with the increase in the period, more light gets reflected off the substrate and is lost.
For INC, as the period increases, the absorption first increases and then slowly starts to decrease, with a similarly opposite profile for its reflectance.
This behavior for the INC is expected, as when the period increases, the spacing between each cone increases, and the reflection on the top is less but more light travels to the space between the structures and strikes the substrate, where the back-reflected light can be re-reflected or absorbed at the sidewall of the tapered structure. However, similar to other structures, with very large spacing, the absorption decreases as light is predominantly lost due to reflection and the net absorptance decreases. The difference in anti-reflective properties between the dense and sparse models is attributed to different effective refractive index mismatches at the air/SC interface \cite{jung2010strong} and enhanced multiple scattering effects in the dense NCs \cite{chattopadhyay2010anti, bai2014one}. \\
\tab Further, it can be seen that the absorptance of the NC structure is better than all the structures for the entire period, and the absorptance of the INC is better than HG for larger periods. In terms of the reflection properties, the INC structure has the highest reflectance, and the NC has the lowest. While for transmittance, the HG structure has the highest, followed by NC.

\begin{figure}[t]
        \centering
        \subfloat[]{\label{fig:na}\includegraphics[width=3.45in]{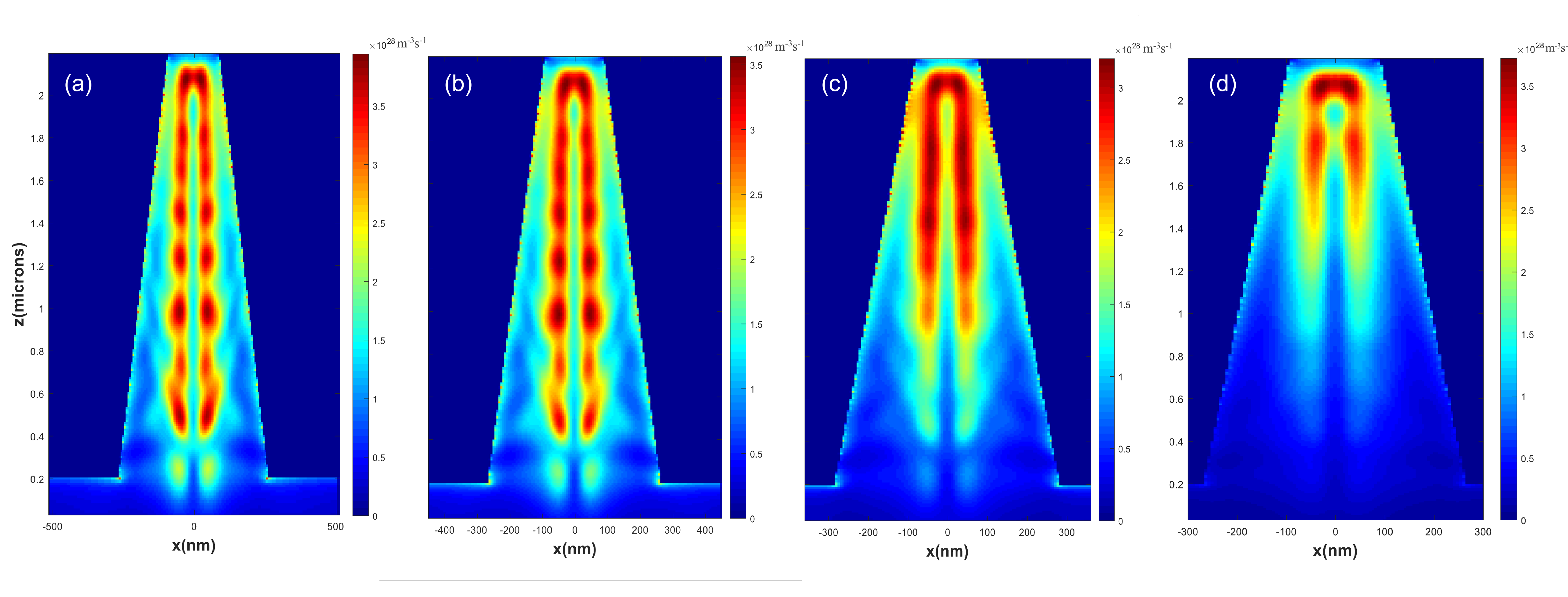}}
        
        \qquad
        \subfloat[]{\label{fig:nb}\includegraphics[width=3.45in]{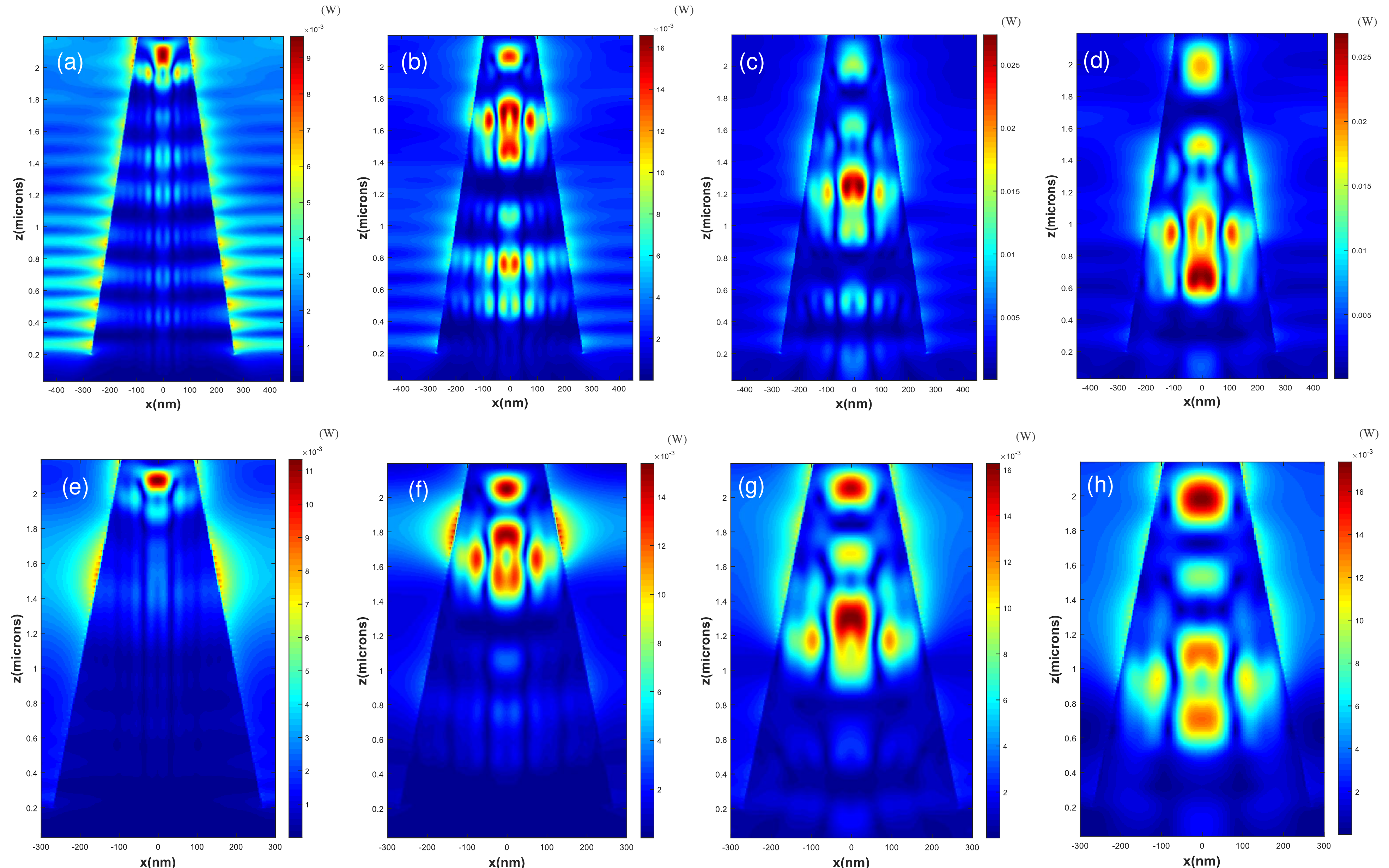}}
        \caption{\textbf{(i)} Photogeneration  profiles  of  the  hourglass solar cell with D/P : (a) 0.35, (b) 0.4, (c) 0.5, (d) 0.6. \textbf{(ii)} The field power profiles across the nanocone for wavelengths (a) 500 nm with D/P = 0.4, (b) 600 nm with D/P = 0.4, (c) 700 nm with D/P = 0.4, (d) 800 nm with D/P = 0.4, (e) 500 nm with D/P = 0.6, (f) 600 nm with D/P = 0.6, (g) 700 nm with D/P = 0.6, (h) 800 nm with D/P = 0.6. }
        \label{fig:n}
\end{figure}

\begin{figure}[t]
        \centering
        \subfloat[]{\label{fig:qb}\includegraphics[width=3.45in]{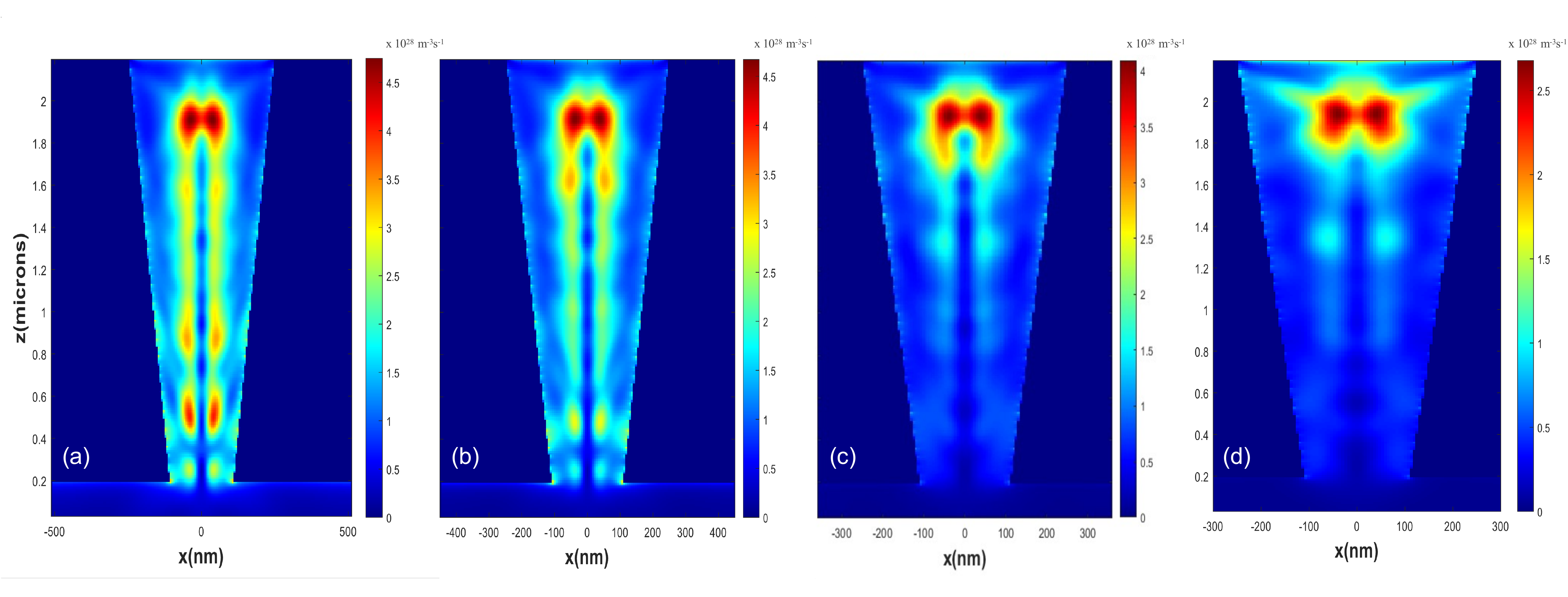}}
        
        \qquad
        \subfloat[]{\label{fig:qa}\includegraphics[width=3.47in]{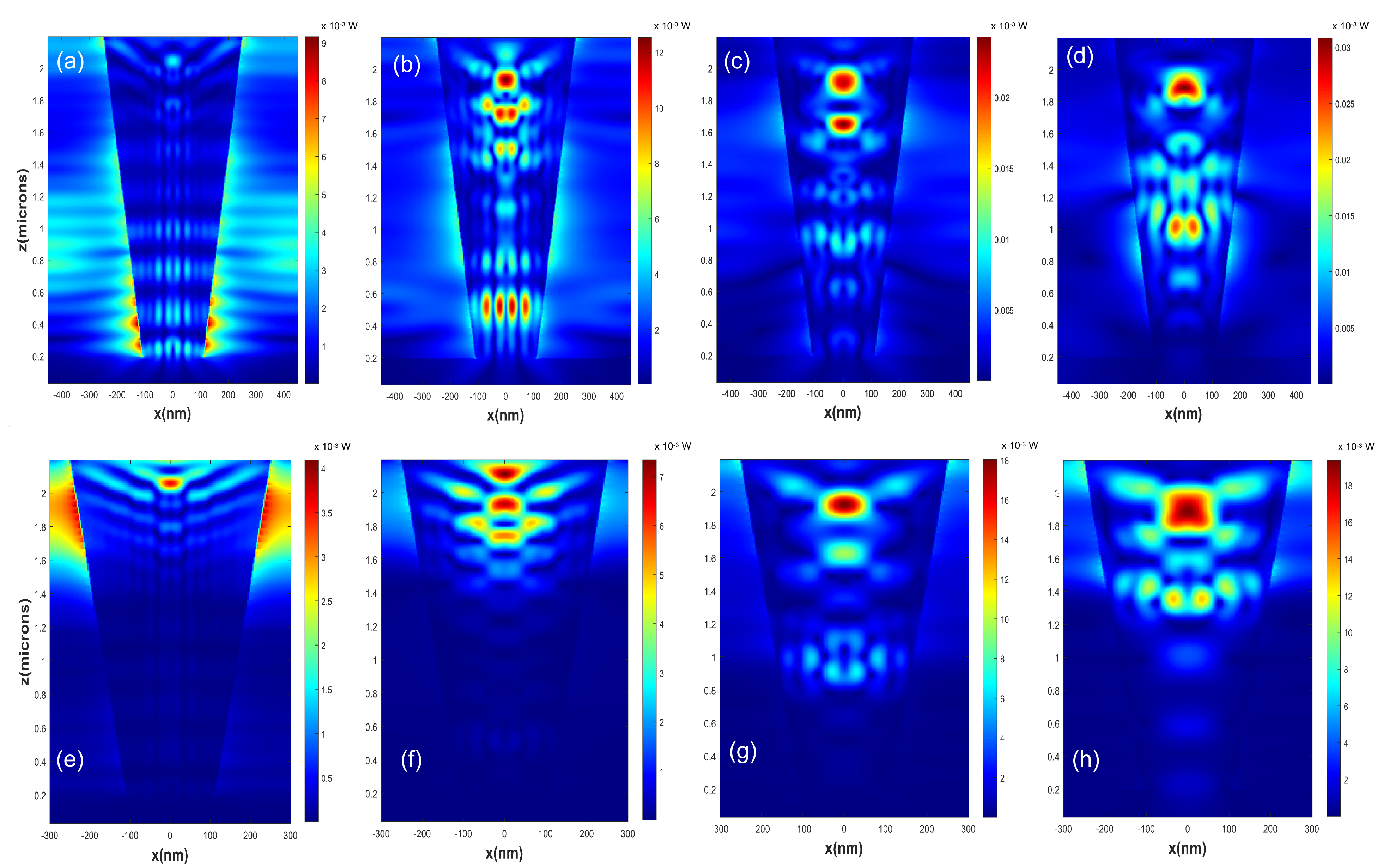}}
        \caption{\textbf{(i)} Photogeneration  profiles  of  the  hourglass solar cell with D/P : (a) 0.35, (b) 0.4, (c) 0.5, (d) 0.6. \textbf{(ii)} The field power profiles across the inverted nanocone for wavelengths (a) 500 nm with D/P = 0.4, (b) 600 nm with D/P = 0.4, (c) 700 nm with D/P = 0.4, (d) 800 nm with D/P = 0.4, (e) 500 nm with D/P = 0.6, (f) 600 nm with D/P = 0.6, (g) 700 nm with D/P = 0.6, (h) 800 nm with D/P = 0.6. }
        \label{fig:p}
\end{figure}

\begin{figure}[t]
        \centering
        \subfloat[]{\label{fig:pb}\includegraphics[width=3.45in]{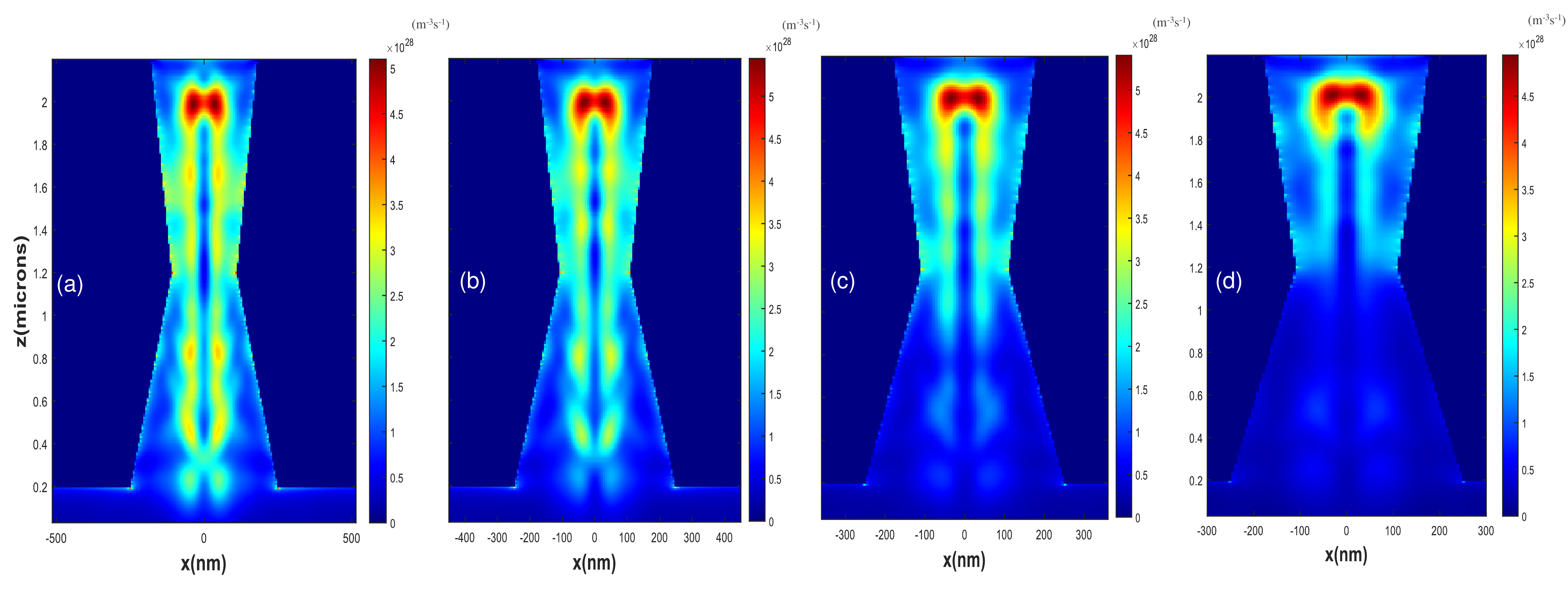}}
        
        \qquad
        \subfloat[]{\label{fig:pa}\includegraphics[width=3.45in]{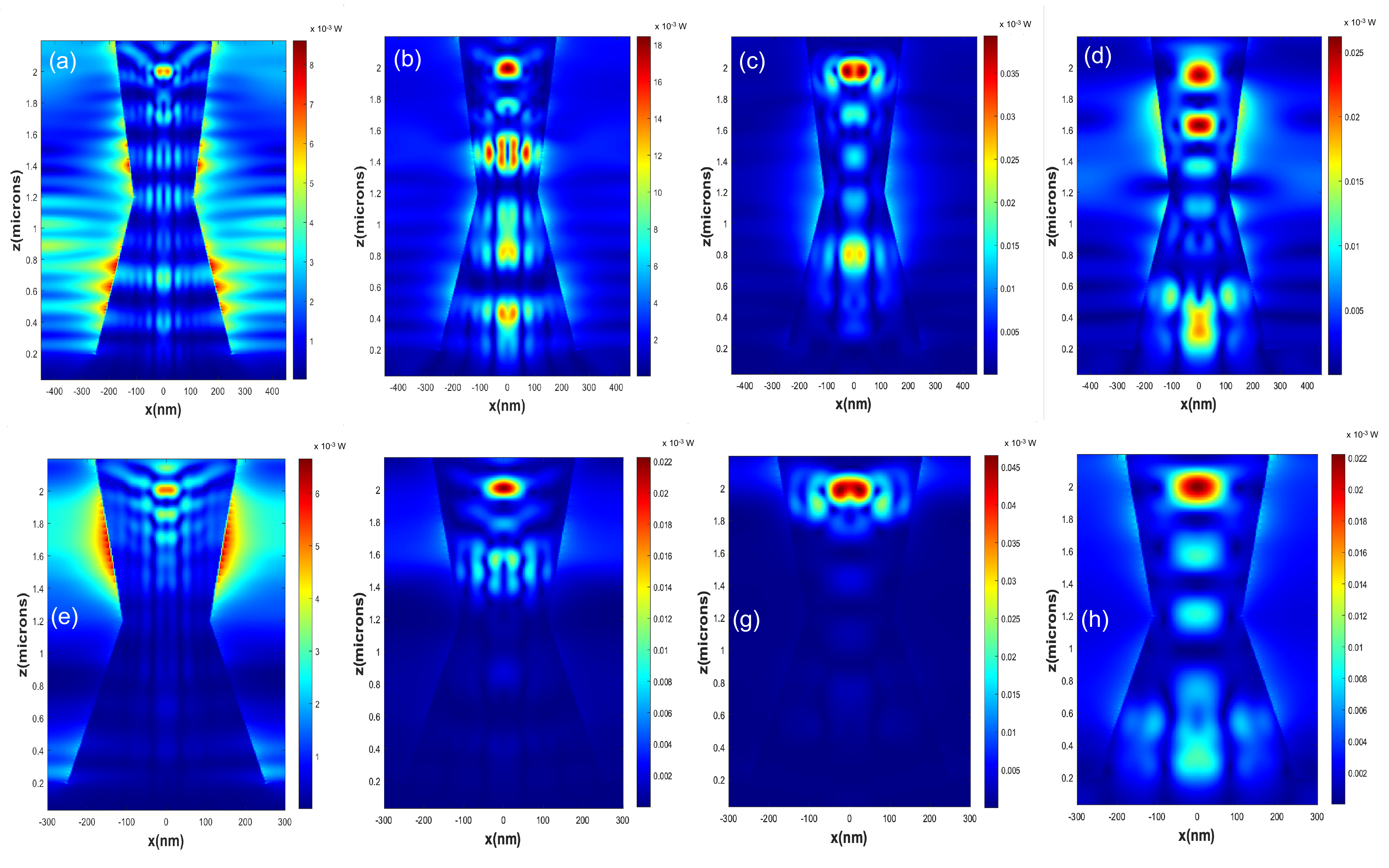}}
        \caption{\textbf{(i)} Photogeneration  profiles  of  the  hourglass solar cell with D/P : (a) 0.35, (b) 0.4, (c) 0.5, (d) 0.6.
       \textbf{(ii)} The field power profiles across the hourglass for wavelengths (a) 500 nm with D/P = 0.4, (b) 600 nm with D/P = 0.4, (c) 700 nm with D/P = 0.4, (d) 800 nm with D/P = 0.4, (e) 500 nm with D/P = 0.6, (f) 600 nm with D/P = 0.6, (g) 700 nm with D/P = 0.6, (h) 800 nm with D/P = 0.6.  }
        \label{fig:q}
\end{figure}

Therefore, in terms of these properties, it seems that a dense structure is expected to be more efficient. However, along with photo-absorption, one more important factor that decides the efficiency of a solar cell is carrier extraction, which depends on the photogeneration across the nanowire and the overlap of the photogeneration with the junction region. From the photogeneration profiles across the NC for different periods in Fig~\ref{fig:na}, we see that as the period increases, the photogeneration spreads out across the entire length of the nanocone. This helps us attain a longer absorption length. Which, in this case, is better than the photogeneration at large nanocone angles, as its obtained without compensating for the loss of absorption due to the thinner top region. In Fig~\ref{fig:nb}, we see the field profiles across the NC for two different periods. It can be seen that when the period is low (D/P is large, dense structure), the field presence in the lower regions of the NC is very less, and the fields are mostly confined towards the top regions as supported by the respective wavelength modes. On the other hand, when the period is large (D/P is small, sparse structure), there is a significant field presence in the lower regions along with the upper regions for all wavelength modes. Thus, we get the extended photogeneration for the sparse structures. Similar profiles can be observed for INC and HG structures as well. In Fig~\ref{fig:p}, \ref{fig:q}, we see the photogeneration and the field profiles for the INC and HG structures respectively, and observe the similarly extended photogeneration and large field presence in the lower parts of the sparse structures having larger periods.

\begin{figure}[t]
  \centering
  \includegraphics[width=3.55 in]{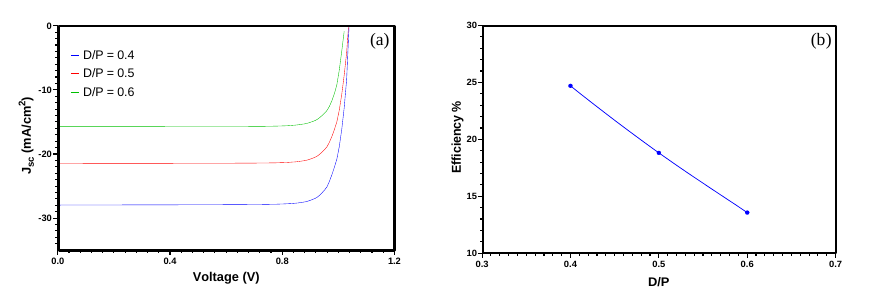}\\
  \caption{(a) IV curves, (b) Conversion efficiency of theGaAs/AlGaAs pin radial junction nanocone for different diameter/period ratios.}
  \label{fig:o}
\end{figure}

\tab This has a great impact on the electrical properties of the radial junction solar cell, as the carrier extraction is in the radial direction. The junction, if properly aligned with this long photogeneration region, can effectively separate the carriers and produce more photocurrent.  A study on observing the same is done by Bai $et$ $al.$ cells \cite{bai2014one} on Si nanocone solar, and by us with GaAs/AlGaAs pin junction radial nanocone solar cell \cite{majumderpv} where it is observed that in contrast to the optical properties, there is a huge increase in the efficiency as the period is increased, i.e., for sparse NCSCs, the highest efficiency is obtained despite the better anti-reflective properties of the dense NCSCs. This change in efficiency comes from the improvement in $J_{sc}$ as seen from Fig~\ref{fig:o} \cite{majumderpv}, resulting from improved photogeneration profile and lower recombination. We can expect a similar result for the case of HG and INC structures. However, the extent of this change for HG and INC structures is subjected to further studies.


\section{Conclusions}
\label{sec:conc}
An optical study of different nanocone, inverted nanocone, and hourglass structures have been done. All the structures show an improved performance than symmetric NWSCs. It is observed that the performance of the nanocone is the best among all the designs for a wide variety of angles in terms of its light-collecting power, and its best performance is obtained at a nanocone angle $\sim$ 7$^{\circ}$. The improved performance of NC structures is attributed to the anti-reflection properties attributed to its sidewall shape and the effective increase in light absorption length across the structure, thereby introducing longer photogeneration profiles. The improved performance of INC structure is significantly observed for nanocone angle $\sim$ 4$^{\circ}$ where the TIR is expected to be maximum. TIR confines light in the inside of the nanocone for a longer time without getting transmitted, thus resulting in very low transmittance and improved absorptance, but it suffers from large reflection from the top, which can be solved by anti-reflection coatings. It can also trap light in the space between the individual cells due to its taper, which can be absorbed later by the structure. HG structure is a combination of both structures and shows a balanced property for the wide range of angles. It has the best tolerance to sidewall angles than the other two structures and has the best performance at an angle $\theta_{b}$ = $\sim$ 9$^{\circ}$, at this point, the top inverted nanocone attains the angle $\theta_{t}$ = $\sim$ 4.5$^{\circ}$ and further increase in angle leads to a loss of TIR. However, its performance is lower than both the structures for a very wide range of wavelengths, but for very small periods, its performance is better than INC structures. The performance of the structures were observed for different sun angles (up to 30$^{\circ}$ from the normal) and it was observed that with small increase in angles, the NC and HG structures show better absorption of light, which can be attributed to their better light trapping properties; whereas, the absorptance of the NC decreases continuously as the sun angle deviates further from the normal. The INC also shows a better tolerance to the changes in sun angles upto 30$^{\circ}$ than the HG and INC structures. The period has a significant impact on the performance of these tapered solar structures. For larger periods (sparse structures), the light is more effectively absorbed across the length of the structures for a wide range of wavelengths which leads to the increase in the effective absorption length and a corresponding increase in photogeneration. This will result in an improved carrier separation and extraction for the radial junction solar cell structures and thus will lead to higher efficiency.



\bibliography{references}
\bibliographystyle{IEEEtran}

\end{document}